\documentclass[twocolumn]{aastex63}

\newcommand{\vx}{\vec{x}}

\received{June 1, xxxx}
\revised{January 10, xxxx}

\submitjournal{ApJ}

\shorttitle{}
\shortauthors{Wang et al.}

\usepackage{amsmath}
\usepackage{subfigure}
\usepackage{graphicx}
\usepackage{lineno}
\usepackage{orcidlink}
\begin{document}
\title{Towards Optimal Reconstruction of Shear Field with PDF-Folding}
\correspondingauthor{JUN ZHANG}
\email{betajzhang@sjtu.edu.cn}

\author{Haoran Wang~\orcidlink{0000-0001-5945-8399}}
\affiliation{Department of Astronomy, Shanghai Jiao Tong University, Shanghai 200240, China}

\author{Jun Zhang~\orcidlink{0000-0003-0002-630X}}
\affiliation{Department of Astronomy, Shanghai Jiao Tong University, Shanghai 200240, China}
\affiliation{Shanghai Key Laboratory for Particle Physics and Cosmology, Shanghai 200240, China}

\author{Hekun Li~\orcidlink{0000-0002-0610-2361}}
\affiliation{Department of Astronomy, Shanghai Jiao Tong University, Shanghai 200240, China}

\author{Cong Liu~\orcidlink{0009-0006-2694-6752}}
\affiliation{Department of Astronomy, Shanghai Jiao Tong University, Shanghai 200240, China}

\begin{abstract}
Weak lensing provides a direct way of mapping the density distribution in the universe. To reconstruct the density field from the shear catalog, an important step is to build the shear field from the shear catalog, which can be quite nontrivial due to the inhomogeneity of the background galaxy distribution and the shape noise. We propose the PDF-Folding method as a statistically optimal way of reconstructing the shear field. It is an extention of the PDF-SYM method, which is previously designed for optimizing the stacked shear signal as well as the shear-shear correlation for the Fourier\_Quad shear estimators. PDF-Folding does not require smoothing kernels as in traditional methods, therefore it suffers less information loss on small scales, and avoids possible biases due to the spatial variation of shear on the scale of the kernel. We show with analytic reasoning as well as numerical examples that the new method can reach the optimal signal-to-noise ratio on the reconstructed shear map under general observing conditions, i.e., with inhomogeneous background densities or masks. We also show the performance of the new method on real data around foreground galaxy clusters.

\end{abstract}


\section{Introduction} \label{sec:intro}
Cosmic shear refers to the coherent shape distortion of the background galaxies due to the gravitational lensing effect by the foreground large scale density fluctuation \citep{Kaiser1992,Bacon2000,Kaiser2000,Wittman2000,Bartelmann2001,Hoekstra2008,Mandelbaum2018}. It has been routinely measured in galaxy surveys for the purpose of not only constraining the cosmological parameters 
\citep{Schrabback2010,Heymans2013,Kilbinger2013,Fu2014,Hikage2019,Heymans2021,Asgari2021,Doux2022}
, but also reconstructing density profiles of the dark matter halos, voids, filaments \citep{Mandelbaum2006,Melchior2014,Clampitt2016,Sanchez2016,Luo2018,Dong2019,Schrabback2021,Xu2021,Fong2022,Wang2022}. 

To our knowledge, lensing is so far the only direct way of mapping out the matter (surface) density field, making it a particularly valuable tool in modern cosmology. The generated mass/density map can provide details about the small scale structure of the universe and the interaction between galaxies, clusters, and the cosmic web. It holds the information about the integrated density fluctuation along the line of sight. Compared with shear two-point correlations, popular applications on the convergence map, such as N-point statistics \citep{Secco2022}, 
peak statistics \citep{Fan_2007,Dietrich2010,Fan2010,Liu_2015,Zorrilla2016,Liu_2016,Shan_2017,Chen_2020,Zhang_2022,Liu_2022,Liu2023}, and Minkowski functionals \citep{Petri2013,Fang2017,Liu2022}, can provide complementary information about non-Gaussian density field at late times generated by non-linear gravitational collapse on small scales. 
It is usually convenient to implement these methods to the density field directly and get the constraint to cosmological parameters and models.
The mass maps can also be intrinsically useful. For example, using the DES Science Verification mass map, \cite{Clerkin2017} shows that the one-point distribution of the density field is more consistent with log-normal than Gaussian. Combining mass maps with the spatial distributions of stellar mass or galaxy clusters enable us to study the relation between the visible baryonic matter and invisible dark matter. Using mass maps to constrain galaxy bias \citep{Chang2016}, the relation between the distribution of galaxies and matter, can in turn aid cosmological probes other than weak lensing. It also enables simple tests for systematic errors in the galaxy shape catalogs.

The surface density field can be constructed from the background shear field through a linear transformation \citep{KS93}. This process is often refined by adopting prior knowledge on the density field through the maximum likelihood method, including Wiener filtering, Sparse priors \citep{Leonard2014,Starck2015,Li2021,Price2021}, DeepMass \citep{Jeffrey2020}. 
However, all the methods focus on deriving the $\kappa$ map from the shear map, but much less on how to generate a shear map from the shear catalog, which is perhaps an equally important problem. Currently, the shear field is typically made by taking the weighted sum of the shear estimators within a given smoothing kernel. The choice of the kernel size can be quite nontrivial: it should be neither too small for keeping enough source galaxies, nor too large for the sake of preserving a reasonably good spatial resolution. A large kernel may also introduce systematic errors in the shear field due to the coupling between the inhomogeneity of the shear field and that of the source density on the kernel scale. In this paper, we aim at solving these problems by proposing a new way to extract the shear map from a shear catalog. Not only that we try to avoid the systematic errors aforementioned, we also hope to reach the optimal statistical uncertainty in the reconstructed shear map. We call this new method PDF-Folding (called PF method hereafter), as it is based on symmetrizing the probability distribution function (PDF) of the shear estimators, similar to the PDF-SYM method  previously proposed in \cite{Zhang2017}.

The rest of the paper is organized as follows.
In \S\ref{method}, we introduce the PF algorithm. In \S\ref{numerical}, we test the accuracy of this method using the mass map computed from the Illutris simulation \citep{Nelson2015}, and demonstrate the accuracy of the PF method in the presence of inhomogeneous source distribution. As a check of the actual performance of PF, we apply the method on the real shear catalog around a few massive foreground galaxy clusters in \S\ref{real}. We give a brief conclusion in \S\ref{conclusion}, and discuss some standing issues in the PF method that the users should be careful with.

\section{Method} 
\label{method}

\subsection{The PDF-SYM Method for Shear Recovery}
PDF-SYM is introduced in \cite{Zhang2017} as a new statistical approach of estimating the shear signal from the shear estimators. It aims at achieving the minimum statistical error (the Cramer-Rao Bound) without introducing systematic biases. Although it is developed based on the Fourier\_Quad shear estimator, the idea is in principle applicable to shear estimators of any form. 

To demonstrate the idea, let us assume that the shear estimator is simply the galaxy ellipticity $e$ of an ideal form, i.e., the measured $e$ is related to the intrinsic ellipticity $e_I$ and the shear signal $g$ via $e=e_I+g$. Note that realistic shear estimators typically require additional corrections, or even take some unconventional forms \citep{Zhang2011a,Sheldon2017}. These changes however does not affect our discussion below, and can be easily incorporated into the PDF-based algorithms, as shown in \cite{Zhang2017}. For simplicity, we have also neglected the sub-indices of the ellipticity and shear. The basic idea of PDF-SYM is to find the value of a pseudo signal $\hat{g}$, which can best symmetrize the PDF of the corrected shear estimator $\hat{e}(=e-\hat{g})$. The new PDF of $\hat{e}$ is related to the PDF of the intrinsic ellipticity $e_I$ via $P(\hat{e})d\hat{e}=P_I(e_I)de_I$, and we have:
\begin{equation}
\label{pi}
P(\hat{e})=P_I(\hat{e}+\hat{g}-g).
\end{equation}
Since $P_I$ is a symmetric function, one can see from eq.(\ref{pi}) that $P(\hat{e})$ is best symmetrized when $\hat{g}$ is equal to the true shear value $g$. The symmetry level of the PDF can be quantified by comparing the galaxy number counts within bins that are symmetrically placed on the two sides of zero. 

The above algorithm is only well defined for recovering a single shear signal. In the following sections, we try to extend it in a way to reconstruct a shear field, which is the main purpose of this work.

\subsection{Reconstruction of a shear field}
\label{rsf}

It is helpful to start with a simple example. Let us assume that the shear field is one dimensional, i.e. $g(x)=ax+b$, in which $x$ is the coordinate in the range of [-1, 1]. For further simplicity, we can assume that the source galaxies are evenly distributed along $x$. A naive idea of applying PDF-SYM would be to adopt the same form of spatial distribution for the pseudo shear signal, i.e., $\hat{g}(x)=\hat{a}x+\hat{b}$, and try to find the best values of $\hat{a}$ and $\hat{b}$ to symmetrize the overall PDF of the galaxies. This is indeed a natural choice for modelling the local distribution of the shear field if the region is small, and is exactly what we do at the early stage of this project. It turns out that in this case, the parameter $\hat{b}$ can be constrained by symmetrizing the PDF, but not the parameter $\hat{a}$. The change of $\hat{a}$ has opposite effects on the ellipticity distributions for galaxies located at $x<0$ and $x>0$, the overall symmetry of the whole PDF is therefore not affected by $\hat{a}$.

\begin{figure}
    \centering
    \includegraphics[width=\linewidth,height=0.5\linewidth]{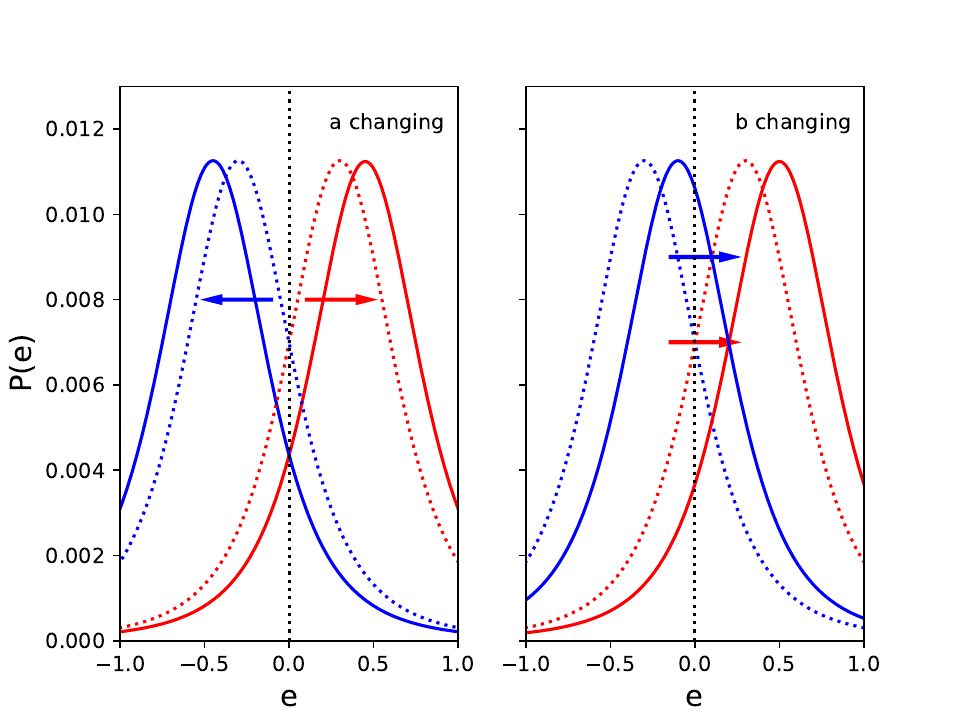}
    \caption{The behavior of the ellipticity PDF of galaxies located at x$<$0 (in blue color) or x$\geq$0 (in red color) when the parameter a (left panel) or b (right panel) changes. The dotted and solid lines correspond to the PDFs before and after the change of the parameter respectively.}
    \label{fig:ex2}
\end{figure}

This is illustrated in fig.\ref{fig:ex2}, in which we show the impact of $\hat{a}$ and $\hat{b}$ on the PDF. For clarity, in the figure, we split the PDF for the galaxies with $x<0$ and $x\ge 0$, shown with the blue and red colors respectively. 
As one can see, the parameter $\hat{b}$ moves the blue and red populations along the same direction, therefore can change the symmetry of the PDF. The parameter $\hat{a}$, on the other hand, moves the two groups of galaxies towards opposite directions, thus does not affect the overall symmetry of the whole PDF.

A possible remedy for this problem is to invert the sign of the corrected shear estimators $\hat{e}$ for galaxies of $x<0$. The symmetry of the resulting new PDF becomes sensitive to $\hat{a}$. 
Its value can therefore be found by symmetrizing the folded PDF. This is why our new method is called PDF-Folding. More generally, for a shear field parameterized by a set of orthogonal functions, we find that one can recover the coefficients one-by-one, and each time by symmetrizing the PDF that is folded at the places where the corresponding basis function changes its sign. 

To realize this idea, let us consider a general shear field $g(\vec{x})$ in 2D. It can be expanded with a set of orthogonal functions: $g(\vec{x})=\sum_k{a_kf_k(\vec{x})}$.
The question becomes how to estimate each mode's magnitude $a_k$. To characterize the PDF of the corrected shear estimator $\hat{e}$, we define $u_i$ ($i=0, \pm 1,...,\pm l$) as the boundaries of the bins placed symmetrically on the two sides of zero, i.e., $u_i=-u_{-i}$. In total there are $2(l+1)$ bins. Note that the outer boundaries of the two outmost bins are at infinite. Assuming we want to recover $a_w$, the estimated shear field is then written as $\hat{g}(\vec{x})=\hat{a}_wf_w(\vec{x})$, in which we caution that the summation sign is not present. $\hat{a}_w$ is the presumed value of $a_w$. The number of galaxies $N_i$ of the $i^{th}$ bin in our new PDF-Folding scheme is defined as:
\begin{eqnarray}
\label{ni}
N_i&=&\bar{n}\int_{i\cdot f_w(\vec{x})\ge 0} d^2\vec{x} \vert f_w(\vec{x}) \vert\int_{u_{\vert i\vert-1}}^{u_{\vert i\vert}}d\hat{e} P(\hat{e})\\ \nonumber
&+&\bar{n}\int_{i\cdot f_w(\vec{x})<0} d^2\vec{x} \vert f_w(\vec{x}) \vert\int_{-u_{\vert i \vert}}^{-u_{\vert i\vert-1}}d\hat{e} P(\hat{e})
\end{eqnarray}

For now, we simply assume that the galaxy number density $\bar{n}$ is a constant. $P$ is the normalized PDF, which we assume does not change with position. The form in eq.(\ref{ni}) is quite different from the usual definition of $N_i$, which is simply $\bar{n}\int d^2\vec{x} \int_{u_{i-1}}^{u_{i}}d\hat{e} P(\hat{e}) $ (for $i>0$). The idea of folding is manifested by mixing the galaxies on the two sides of zero in PDF according to the sign of $f_w(\vec{x})$ (which is known) in eq.(\ref{ni}). The factor $\vert f_w(\vec{x}) \vert$ is an additional weight for filtering out the contamination from other orthogonal modes, as we show next.

To find out how the value of $\hat{a}_w$ can change the symmetry of the PDF, let us calculate $N_i-N_{-i}$, which is:
\begin{eqnarray}
\label{ni3}
&&N_{i (>0)}-N_{-i} \\ \nonumber
&=&\bar{n}\int d^2\vec{x} f_w(\vec{x})\left[\int_{u_{i-1}}^{u_{i}}-\int_{-u_i}^{-u_{i-1}}\right]d\hat{e} P(\hat{e})
\end{eqnarray}
Note that the integration without lower and upper bounds means integrating over the whole available range of the variable. Using eq.(\ref{pi}), we can relate the function $P$ with the intrinsic PDF $P_I$ (which is symmetric) as: 
\begin{eqnarray}
\label{pi2}
&&P(\hat{e})=P_I(\hat{e}+\hat{g}-g) \\ \nonumber
&=&P_I\left[\hat{e}+\hat{a}_wf_w(\vec{x})-\sum_k{a_kf_k(\vec{x})}\right]\\ \nonumber
&\approx&P_I(\hat{e})+\left[\hat{a}_wf_w(\vec{x})-\sum_k{a_kf_k(\vec{x})}\right]P_I'(\hat{e})
\end{eqnarray}
where $P_I'(e)=dP_I(e)/de$.
The last step above is from Taylor's expansion to the first order, as the shear signal is assumed to be small. Using the results in eq.(\ref{pi2}), we can get:
\begin{eqnarray}
\label{ni4}
&&\left[\int_{u_{i-1}}^{u_{i}}-\int_{-u_i}^{-u_{i-1}}\right]d\hat{e} P(\hat{e}) \\ \nonumber
&\approx&2\left[\hat{a}_wf_w(\vx)-\sum_k{a_kf_k(\vx)}\right]\left[P_I(u_i)-P_I(u_{i-1})\right]
\end{eqnarray}
Using the result of eq.(\ref{ni4}) in eq.(\ref{ni3}), and the orthogonality of $f_k(\vx)$, we get:
\begin{eqnarray}
\label{ni5}
&&N_{i (>0)}-N_{-i} \\ \nonumber
&\approx&2\bar{n}(\hat{a}_w-a_w)\left[P_I(u_i)-P_I(u_{i-1})\right]\int d^2\vec{x} f_w^2(\vec{x})
\end{eqnarray}
It is clear that the folded PDF is best symmetrized when $\hat{a}_w=a_w$. This is true for all the bin pairs. Therefore, to estimate $\hat{a}_w$, we just need to minimize the following $\chi^2$\footnote{The definition of $\chi^2$ follows the form given in \cite{Zhang2017}. In our case, the quantity $N_{i}-N_{-i}$ is a number to quantify the symmetry level of the PDF, as required by our method. The statistical mean of $N_{i}-N_{-i}$ is zero when the PDF is fully symmetric with respect to zero. The denominator in our $\chi^2$ definition is the variance of $N_{i}-N_{-i}$.  The minimum of $\chi^2$ indicates that the PDF reaches a best symmetrized state.}:
\begin{equation}
\chi^2=\frac{1}{2}\sum_{i(>0)} \frac{(N_i-N_{-i})^2}{\langle(N_i-N_{-i})^2\rangle}
\end{equation}
The above procedure is repeated for each $\hat{a}_w$ to recover the whole shear field. This is the basic implementation of the PF method. 

For inhomogeneous galaxy distribution, it is not hard to check that a similar conclusion can be reached if we simply replace the weighting of the galaxy number in eq.(\ref{ni}) from $\vert f_w(\vec{x}) \vert$ to $\vert f_w(\vec{x}) \vert/n(\vec{x})$, where $n(\vec{x})$ is the galaxy number density. The form of eq.(\ref{ni5}) remains the same, but without the factor $\bar{n}$. This, however, turns out not to be the end of the story. The formulation can be further improved to achieve two more goals: 1. proper treatment of masks; 2. minimum statistical noise. This is what we discuss next.

\subsection{Optimal Reconstruction Method}
\label{orm}

If we chose the weight to be $\vert f_w(\vec{x}) \vert/n(\vec{x})$, one can immediately identify a problem: what if there are masked areas? A simple option would be to skip the masked areas. However, in this case, the orthogonality of the functions $f_k(\vec{x})$ would not lead to eq.(\ref{ni5}) because of the incomplete domain of integration. Another option would be to use a set of orthogonal functions that are defined in the domain excluding the masked areas. This would be a fine choice, except that the form of $f_k(\vec{x})$ could be quite complicated.

In this work, we take another route: we fill out the masked area with galaxies of a certain number density, e.g., the average number density of the whole field. Each such galaxy is given a random shear estimator/ellipticity, i.e., the shear field is assumed to be zero in the masked areas. This procedure guarantees the completeness of the domain of integration, and therefore the validity of the PF method. 

More generally, we are interested in finding the optimal way of extracting the shear field information from the shear catalog in the case of inhomogeneous galaxy distribution. To do so, we find that the following generalized form of series expansion is useful:
\begin{equation}
\label{formula_alpha}
    g(\vec{x})n^{\alpha}(\vec{x})=\sum_k a_kf_k(\vec{x})
\end{equation}
For recovering the coefficient $a_w$, the galaxy weight and therefore the definitions of the PDF should be updated accordingly as:
\begin{eqnarray}
\label{ni6}
N_i&=&\int_{i*f_w(\vec{x})\ge 0} d^2\vec{x} \vert f_w(\vec{x}) \vert n^{\alpha}(\vec{x})\int_{u_{\vert i\vert-1}}^{u_{\vert i\vert}}d\hat{e} P(\hat{e})\\ \nonumber
&+&\int_{i*f_w(\vec{x})<0} d^2\vec{x} \vert f_w(\vec{x}) \vert n^{\alpha}(\vec{x})\int_{-u_{\vert i\vert}}^{-u_{\vert i\vert-1}}d\hat{e} P(\hat{e})
\end{eqnarray}
In this case, one can show that the difference between the opposite bins is given by:
\begin{eqnarray}
\label{ni50}
&&N_{i (>0)}-N_{-i} \\ \nonumber
&\approx&2(\hat{a}_w-a_w)\left[P_I(u_i)-P_I(u_{i-1})\right]\int d^2\vec{x} f_w^2(\vec{x})
\end{eqnarray}

Note that the galaxy weight is $\vert f_w(\vec{x}) \vert n^{\alpha-1}(\vec{x})$ at $\vec{x}$. To recover $a_w$, e.g., we assume that $\hat{g}(\vec{x})$ is given by $\hat{g}(\vec{x})=n^{-\alpha}(\vec{x})\hat{a}_wf_w(\vec{x})$. Following similar calculations as in \S\ref{rsf}, one can show that eq.(\ref{ni5}) still holds (without the factor $\bar{n}$). Apparently, we now have the exponent $\alpha$ as an additional degree of freedom in our formalism. It turns out that when $\alpha=1/2$, the statistical uncertainty of $a_w$ reaches its minimum. It is therefore our best choice for the shear field reconstruction. We show the details of our proof in Appendix A. 

The above discussion is all about reconstructing the shear field at a given background redshift. More generally, we should consider the redshift distribution of the source galaxies. For the single thin lens case (which is what we consider in this work), the shear fields at different redshifts are related through a simple rescaling with the corresponding critical surface densities. The problem is therefore still two dimensional, and we only need to choose a certain background redshift as a reference. The optimization of the PF method in this case are given in Appendix B.  

In the rest of the paper, we demonstrate some advantages of the PF method using numerical examples, and show some shear/density field reconstruction examples from the real data.

\section{Numerical Examples}\label{numerical}

\begin{figure}[htbp]
\centering
\includegraphics[scale=0.5]{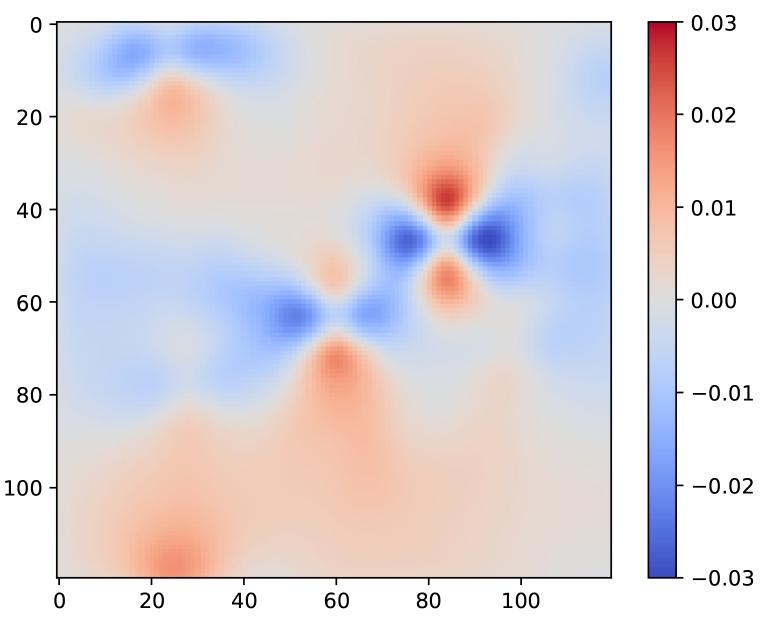}
\caption{Shear field of $g_1$ from the density map of the Illutris-1-Dark simulation within a $6\times6 ({\rm Mpc/h})^2$ comoving area at redshift 0.2. The source redshift is assumed to be at 0.5.}
\label{f_field_ill}
\end{figure}

In this section, we demonstrate several advantages of the PF method with data from the Illutris-1-Dark simulation \citep{Nelson2015}. The side length of the simulation box is $75 {\rm Mpc/h}$ (comoving). We use the snapshot with ID 120 corresponding to $z=0.2$. We select a $6\times6 ({\rm Mpc/h})^2$ (comoving) part containing cluster-like structure, 
and calculate the projected surface density by integrating over the whole box size. The shear field is then deduced from the mass distribution by assuming the source galaxies are all at $z=0.50$, and placed on a $120\times 120$ grid as shown in fig.\ref{f_field_ill} for the $g_1$ component.
This shear field is applied onto a large number of background galaxies whose intrinsic ellipticities are generated according to \cite{Miller2013}. The source galaxies are randomly placed inside the foreground area. To reduce the shape noise, we group every $5\times5$ grid area to form a $24\times 24$ shear map. 

As our first test, we adopt three strategies to reconstruct the shear map from the shear catalog: 1. by taking the local averages of the shear estimators (called Local\_AVE hereafter); 2. by using the PDF-SYM method, which is designed for optimizing the stacked shear signal on the local ensemble of the shear estimators (called Local\_PDF hereafter); 3. by the PF method. In the last method, we expand the shear field with the Fourier series.

\begin{figure}[htbp]
\centering
\includegraphics[scale=0.7]{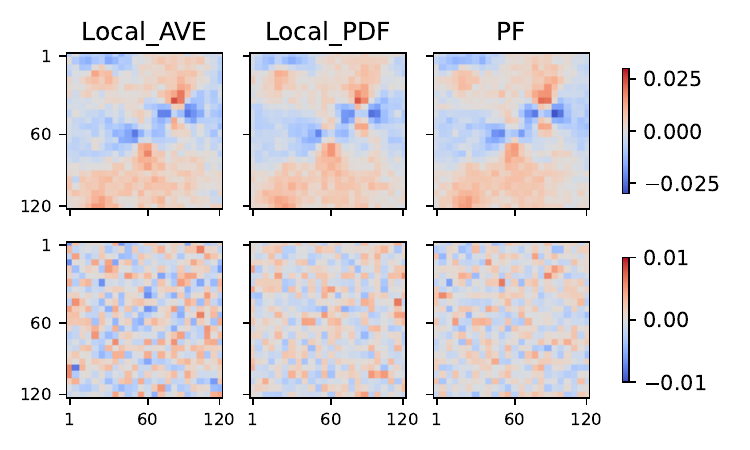}
\caption{From the left to right, we show the recovered shear maps of $g_1$ from the Local\_AVE, Local\_PDF, and PF methods respectively. The upper panels show the shear fields, and the lower panels show the deviations of the results from the input shear map accordingly. }
\label{f_field}
\end{figure}

In our first test, the shear field is applied onto $10^{7}$ background galaxies whose ellipticities follow the form of disk-dominated galaxies. We set the parameter $e_{max}$ in \cite{Miller2013} to be $1.0$.
The recovered shear maps ($g_1$) with the methods of Local\_AVE, Local\_PDF, and PF are shown in the upper panel of fig.\ref{f_field}. The corresponding residuals are shown in the lower panels accordingly. One can see that the PDF-based methods generally generate somewhat less noise than the Local\_AVE method. The advantage would be more obvious if the PDF of the ellipticities has a more significant extension to large values. 

Note that although in this simple example, the Local\_PDF method seems to work similarly well as PF, its performance is actually more sensitive to the background galaxy number density, i.e., there should be enough galaxies in each grid cell, as we show in the real data examples in \S\ref{real}. The PF method avoids this issue by using the background galaxies in the field altogether to form the PDF. It therefore does not require a large smoothing kernel or grid size, and can keep information on smaller scales in principle. 

\begin{figure}
\centering
\includegraphics[scale=0.6]{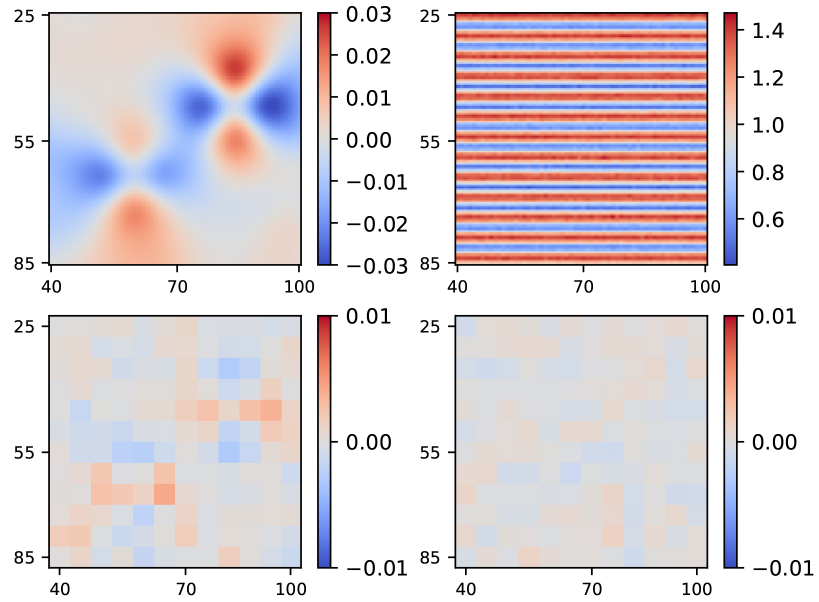}
\caption{Upper left: the central region of the input shear field from Illutris; upper right: the distribution of the galaxy number density field; lower left: the difference between the input shear field and the recovered one from the Local\_PDF method; lower right: the difference between the input shear field and the recovered one from the PF method.}
\label{f_diff_sinmap_compare}
\end{figure}

\begin{figure}
\centering
\includegraphics[scale=0.6]{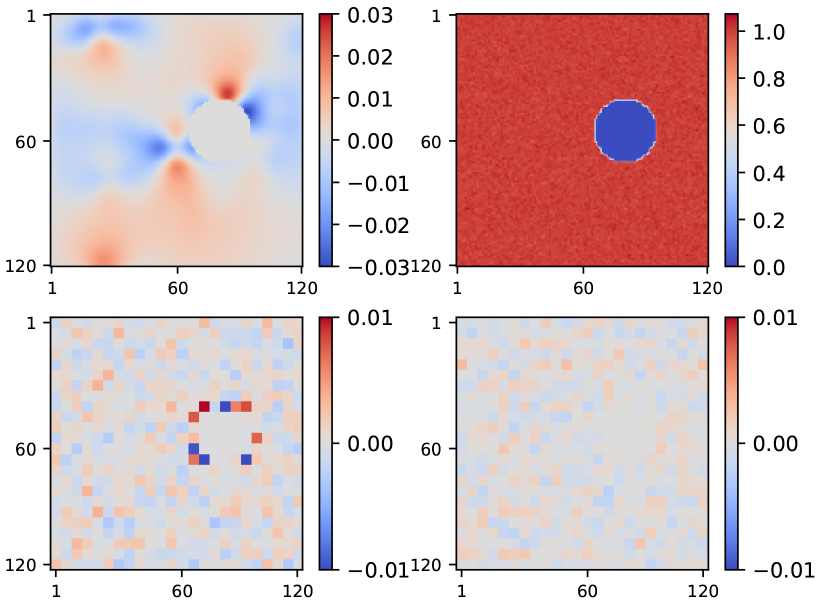}
\caption{The organization of the panels is the same as that of fig.\ref{f_diff_sinmap_compare}. It demonstrates the mask effect on the recovery of the shear field.}
\label{f_mask_patch_compare}
\end{figure}

\subsection{Effect of Inhomogeneous Galaxy Distribution}
\label{inhomo}
The distribution of the background galaxies are often inhomogeneous (as shown in some examples of \S\ref{real}). In traditional methods, if both the shear signal and the background density varies significantly within the smoothing kernel, we expect the coupling of these two types of fluctuations to induce systematic errors in the recovered shear field. In the new PF method, we find that such a bias can be largely avoided. 

To study this effect, we let the background galaxy density vary with the vertical axis y as '$\sin(ky) +const$'  where $k=2\pi/5$, whose period equals the grid size of the recovered shear field. To enhance the signal-to-noise ratio of the systematic bias, we use $5\times10^7$ galaxies with ellipticities set in the range of $[0.05, 0.1]$, 90\% of which follow the ellipticity distribution of the disk-dominated galaxies, and the other 10\% follow that of the bulge-dominated ones. In fig.\ref{f_diff_sinmap_compare}, we show the original shear field in the upper-left panel, the background galaxy distribution in the upper-right panel (the unit of which is arbitrary). The lower two panels show the differences between the recovered shear field and the original one for the Local\_PDF method (left) and the PF method (right). One can see that the shear residuals of the Local\_PDF method are quite significant, and correlated with the signal. In contrast, the PF method performs quite well in the presence of the background inhomogeneity. Note that in the current study, the smoothing kernel for the Local\_PDF is simply the top-hat function within the grid size (of the recovered shear field). If a larger kernel is chosen, the effect is more obvious. For the PF method, we use the expansion defined in eq.(\ref{formula_alpha}) with $\alpha=1/2$. The local number density around each galaxy is calculated by averaging the galaxy numbers within its neighboring $3\times 3$ finer grid cells (the grid on the original $120\times 120$ map). 



\subsection{Effect of Mask}

Masks are just special cases of background inhomogeneity. For similar reasons, the Local\_PDF operation can also generate systematic biases near masks. In this test, we assign a masked area in the central region of the simulated shear map from Illutris, as shown in the upper-left panel of fig.\ref{f_mask_patch_compare}. It can also be seen from the upper-right panel of the same figure, in which we show the distribution of the background galaxy density (the unit is again arbitrary). We apply the shear field to $10^{7}$ background galaxies in total, the ellipticities of which are generated in the same way as those in \S\ref{inhomo}. The background galaxies are evenly distributed in the whole area except the masked region. In the lower panels of fig.\ref{f_mask_patch_compare}, we again show the performances of the Local\_PDF method (left) and the PF method (right) in terms of the differences between their recovered shear fields and the original one. We can see that there are significant residuals near the masked area in the Local\_PDF method. As a comparison, the shear field from the PF method shows no such residuals, demonstrating a consistently better performance. We note that to use PF in this case, one needs to add fake galaxies of random (or zero) ellipticities in the masked area. The number density of the fake galaxies inside the mask is chosen to be comparable to the mean of the whole field.

\subsection{Step-by-step Procedure of the PF Method}
\label{SBS}

Here for simplicity, we still use the ellipticity $e$ to represent the shear estimator, and neglect its sub-index. The application of the PF method for reconstructing the shear field follows the following steps :

\quad 1. Set up a rectangular grid for the shear field; 

\quad 2. Count the galaxy number density $n(\vec{x})$ in each grid;

\quad 3. Fill up the masked area with galaxies of the average number density and random ellipticities/shear estimators;


\quad 4. Determine a set of orthogonal and complete functions, e.g., Fourier series $f_w(x)$,  to parameterize the shear field as: $g(\vec{x})\sqrt{n(\vec{x})}=\sum_w{a_wf_w(\vec{x})}$, where $w$ is the index of each Fourier mode.  
The rest of the steps are about determining the coefficients $a_w$ one-by-one.

\quad 5. Each $a_w$ is determined by changing its assumed value $\hat{a}_w$, so that the PDF of the whole background galaxy sample is best symmetrized. The pseudo shear signal for a galaxy is given by $\hat{g}(\vec{x})=\hat{a}_wf_w(\vec{x})/\sqrt{n(\vec{x})}$ (without summing over all possible values of $w$). For each galaxy, its weight is $\vert f_w(\vec{x}) \vert /\sqrt{n(\vec{x})}$ as defined in eq.\ref{ni6} with $\alpha=1/2$. The sign of $\hat{e}(=e-\hat{g})$ needs to be inverted wherever $f_w(\vec{x})<0$;

\quad 6. Repeat step 5 to find the values of all $a_w$. Calculate the resulting shear field with $g(\vec{x})=\sum_w{a_wf_w(\vec{x})}/\sqrt{n(\vec{x})}$. The resulting shear field can be converted to the $\kappa$ field through, e.g., the K-S inversion algorithm.

Note that in the PF method, the choice of the pixel size is rather arbitrary. Large pixel size is fine for those who are only interested in the large scale behavior of the shear field. On the other hand, if the pixel size is small, say each pixel only containing a handful of source galaxies on average, the method still works fine. In this case, although the recovered shear field in each pixel would become more noisy, the signal-to-noise ratio of the large scale modes would not be affected. 

\section{Application in real data}
\label{real}

In this section, we demonstrate the performance of the PF method with real data. We adopt the shear estimator of the form defined in the Fourier\_Quad shear measurement method \citep{Zhang2015,Zhang2010}, which has been shown to work well with the PDF-SYM method \citep{Zhang2017}. Note that although the Fourier\_Quad shear estimator takes an unconventional form, it works equally well as the ideal shear estimators discussed in \S\ref{method}, as shown in \cite{Zhang2017}. In other words, the whole discussion in \S\ref{method} can be almost trivially replicated for the Fourier\_Quad shear estimator. 

Our shear catalog comes from the HSC galaxy survey Data Release 2 \citep{Aihara2019}. The images are processed by the Fourier\_Quad pipeline in five bands (grizy) individually. The Fourier\_Quad pipeline is previously used to process the imaging data of CFHTLenS \citep{Zhang2019} and DECaLS \citep{Zhang_2022}, and the results of which have all passed the accuracy test using the field-distortion effect. We find that the same pipeline also performs quite well for the HSC data. The details of the measurement are presented in a separate paper (Liu et al., in preparation). In this work, we use the shear catalog of the 'rizy' four bands. Note that in Fourier\_Quad, shear estimators from different exposures or bands are treated as independent ones even if they are from the same galaxy. 

The foreground lens in our examples are three massive galaxy clusters chosen from the SDSS group catalog \citep{Yang2007sdss,Yang2008sdss,Yang2009sdss}. Their information is shown in Table \ref{cluster3}, including their angular positions, redshift ($z_{lens}$), and mass (from the catalog). To reconstruct the surface density map, we first build up the background shear field at a reference redshift ($z_{lens}+0.3$) following the more general version of the PF method (taking into account the background redshift distribution) in Appendix B, in which a modified version of the step-by-step procedures given in \S\ref{SBS} is also provided. The convergence/density map is then converted from the shear map through the standard procedures given in \cite{KS93}. One complication is that what we actually measure from the background galaxy shapes are not the shear, but the reduced shear \citep{Zhang2011}. To improve the accuracy of the recovered density map, one therefore needs to modify the shear estimators with the factor $1-\kappa$, where $\kappa$ is the convergence just derived from the shear map. Usually with several iterations, the shear map and $\kappa$ map can both achieve stable solutions \citep{Liu2014}.
To avoid possible contamination from the cluster members, we only use the background galaxies with redshifts larger than $z_{lens}+0.1$.  


\begin{table}
\centering
\begin{tabular}{|c|c|c|c|} 
\hline
Cluster ID & (RA, Dec) & redshift & mass ($M_{\odot}$)\\
\hline\hline
Cluster A & (197.873,-1.348) & 0.189 & $10^{15.13}$ \\ 
\hline
Cluster B & (355.276,-0.003) & 0.190 & $10^{14.21}$ \\ 
\hline
Cluster C & (210.847,0.128) & 0.184 & $10^{14.47}$ \\ 
\hline
\end{tabular}
\caption{The information of the three clusters used as foreground lenses in \S\ref{real}.}
\label{cluster3}
\end{table}

\begin{figure*}
    \centering
    \includegraphics[scale=0.45]{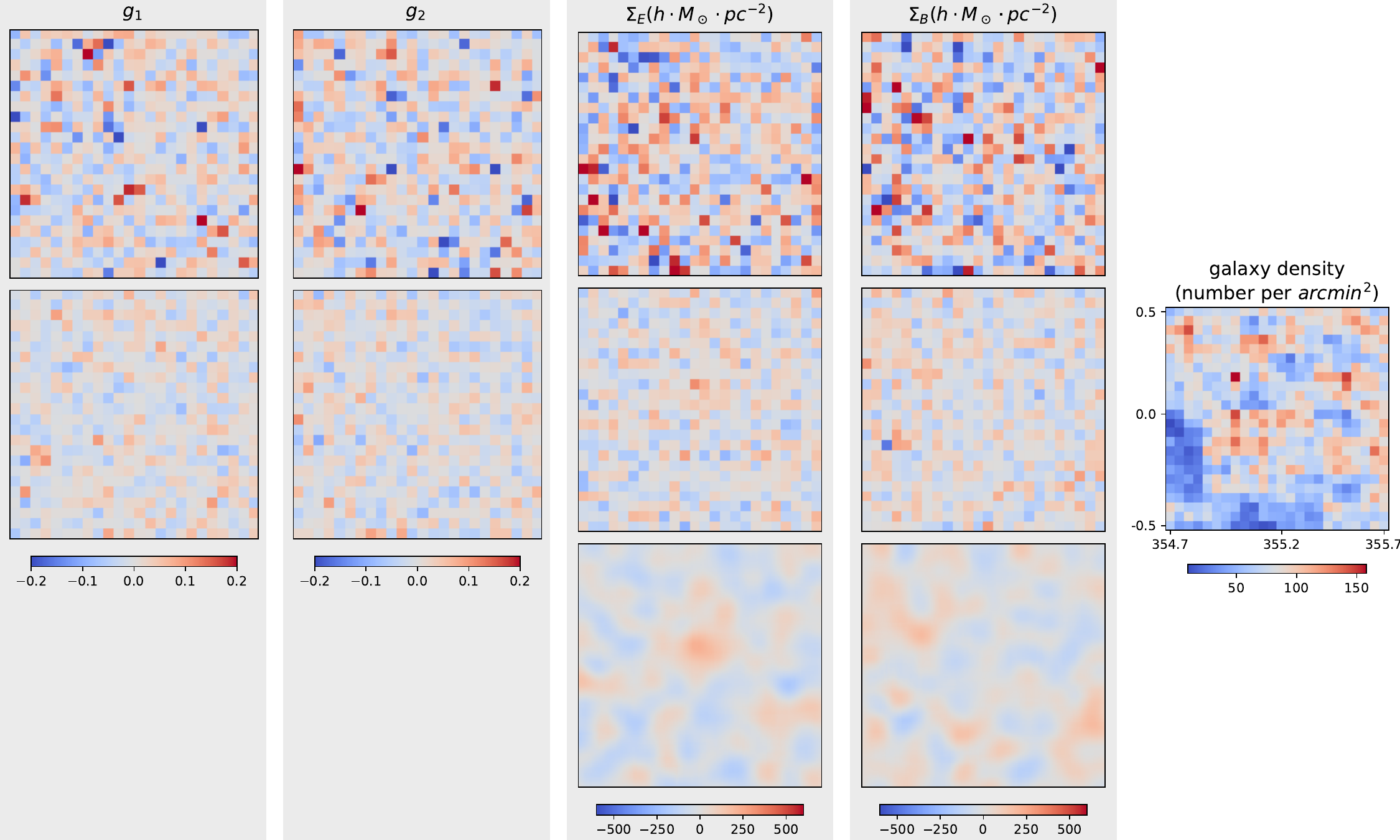}
    \caption{The recovered shear and density maps around Cluster A listed in Table \ref{cluster3}. 
    The first two columns show the $g_1$ and $g_2$ fields for the Local\_PDF method(the first row), Local\_AVE method(the second row) and the PF method (the third row)}. The third and fourth columns show the corresponding surface density fields of E ($\Sigma_E$) and B ($\Sigma_B$) modes respectively. The rightmost column shows the distribution of the background galaxy number density in the same area of the sky. The fourth row shows the density fields recovered also by the PF method, but with a higher spatial resolution.
    \label{f_355}
\end{figure*}
\begin{figure*}
    \centering
    \includegraphics[scale=0.45]{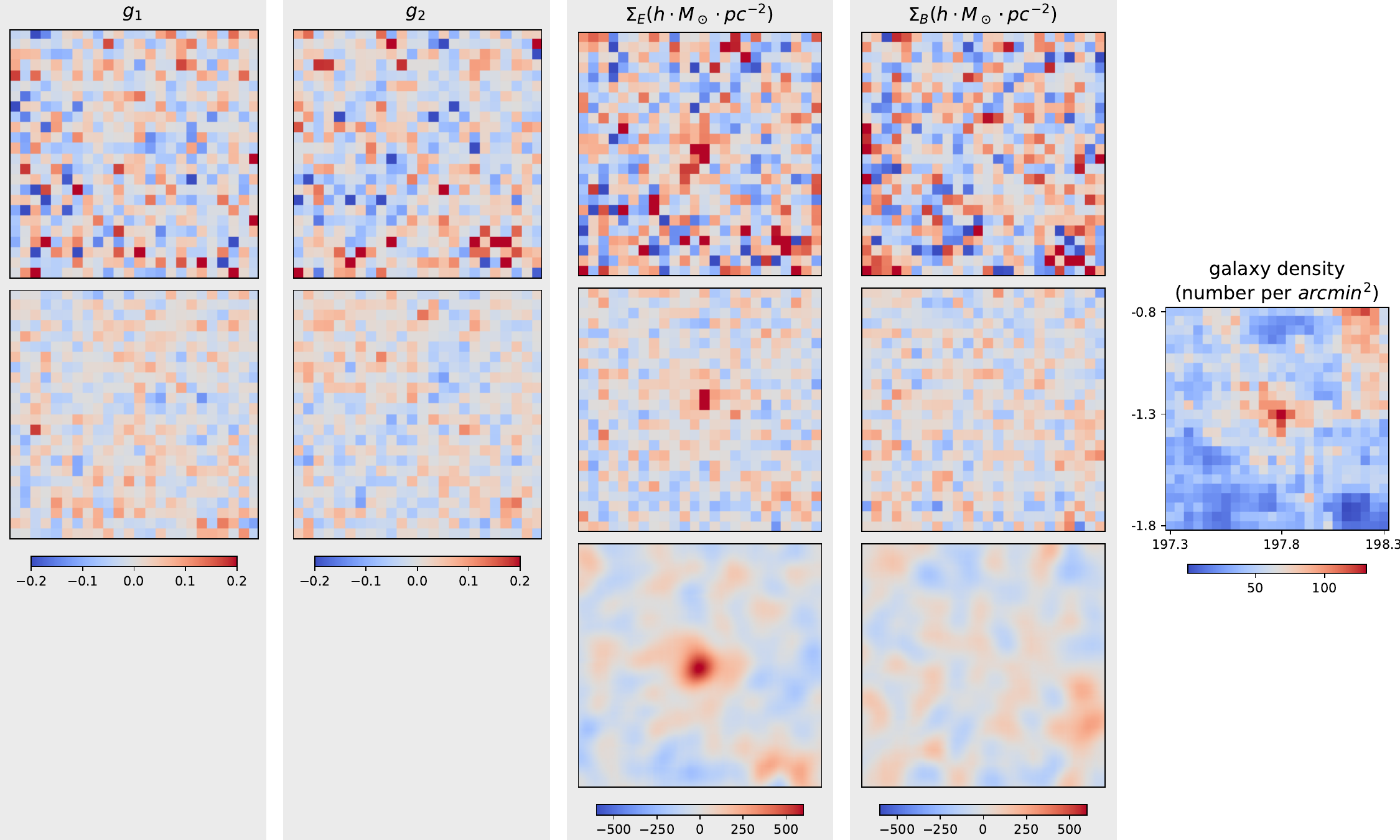}
    \caption{Same as fig.\ref{f_355}, but for Cluster B listed in Table \ref{cluster3}. }
    \label{f_197}
\end{figure*}
\begin{figure*}
    \centering
    \includegraphics[scale=0.45]{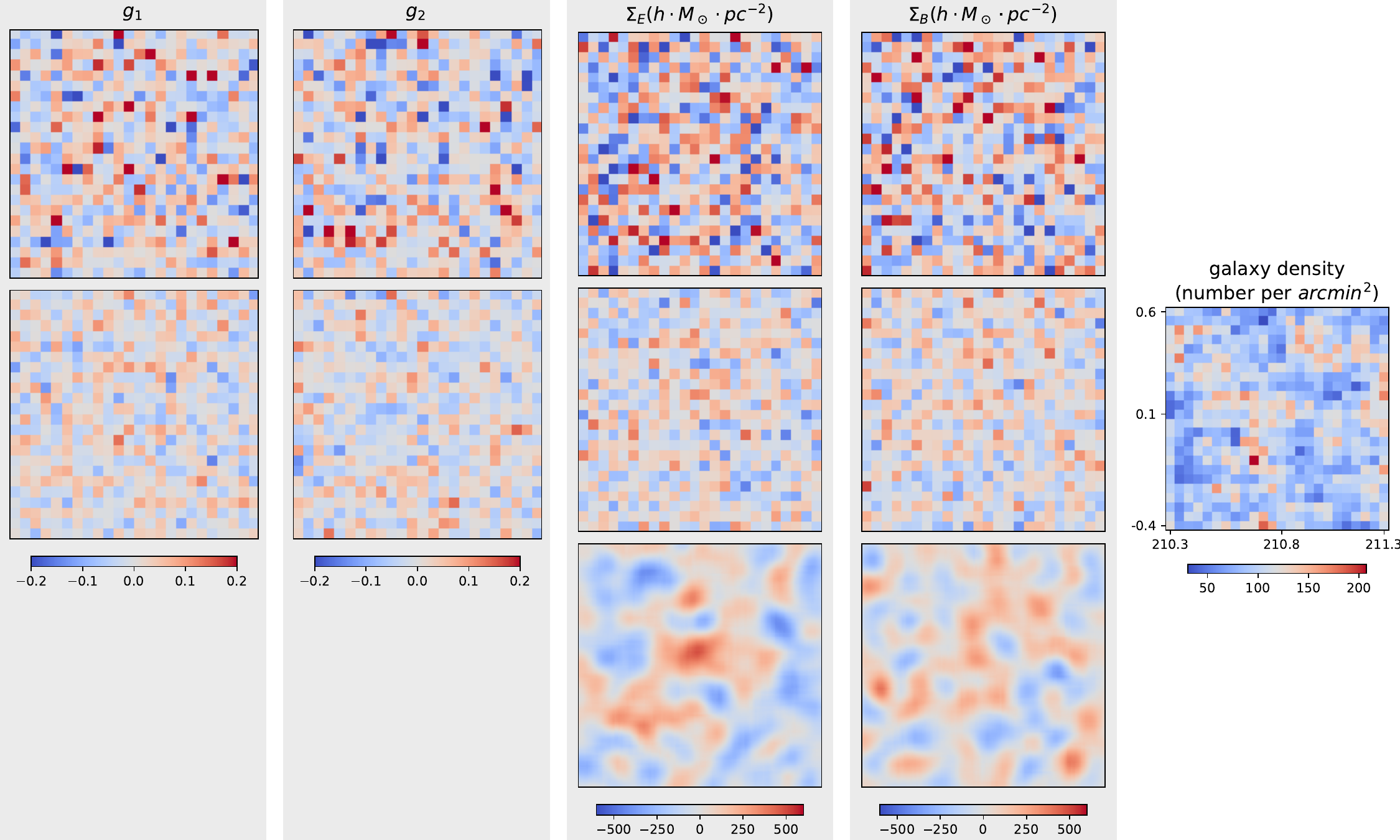}
    \caption{Same as fig.\ref{f_355}, but for Cluster C listed in Table \ref{cluster3}.}
    \label{f_210}
\end{figure*}

For comparison, the Local\_PDF method, Local\_AVE method and the PF method are implemented. The results are shown in fig.\ref{f_355},\ref{f_197},\ref{f_210} for cluster A, B, and C respectively. All the panels in the figures have the dimension of $1^{\circ}\times 1^{\circ}$. In the grey area of each figure, we show the results of Local\_PDF, Local\_AVE and PF in the first three rows respectively, with grid size of $2.5'\times 2.5'$.

We note that in the Local\_AVE method, we use the ensemble average of the local Fourier\_Quad shear estimators to evaluate the local shear field $g_1$ and $g_2$, as shown in eq.(28) of \cite{Zhang2017}. This approach is usually not encouraged to use for the Fouier\_Quad shear estimators because it uses unnormalized quadruple moments of the galaxy, and therefore variations in galaxy luminosity/size can lead to large scatters in shear recovery (this fact however does not cause any problem in PDF-based methods like Local\_PDF and PF here). Here for the purpose of comparison, we still present the results of Local\_AVE, but excluding 8\% highest and 8\% lowest shear estimators to reduce the noise. We would not recommend doing so in practice, as the selection would inevitably introduce local shear bias.


The two left columns show the recovered shear fields $g_1$ and $g_2$ of the two methods. The third and fourth columns show the surface density maps from the E ($\Sigma_E$) and B ($\Sigma_B$) modes of the convergence fields respectively.
The rightmost panel (out of the grey region) shows the background galaxy image density used in the same area of the sky. It is clear from the figures that the results of the PF method are generally less noisy than those of Local\_PDF. Some of the extreme values on the shear maps of Local\_PDF are caused by two reasons: 1. lack of background galaxies; 2. outliers in shear catalog. These two factors fortunately do not affect the performance of the PF method. We consider this feature a significant advantage of PF. It allows us to further increase the spatial resolution of the density map. For example, in the bottom rows of fig.\ref{f_355}, \ref{f_197}, and \ref{f_210}, we show the reconstructed density maps using PF with a much finer grid size of $0.5'\times 0.5'$. The final results are smoothed with a Gaussian filter of $\sigma=1.65'$ to increase the visibility of the density structure. In this case, one can see that the central overdensities of the clusters stand out more clearly. In contrast, we find that it is not feasible to further increase the spatial resolution in the method of Local\_PDF.


\section{Conclusion and Discussions}
\label{conclusion}

Weak lensing provides a direct way of reconstructing the foreground density distribution. So far, The existing algorithms mostly focus on refining the conversion from the shear field to the density map. The construction of the shear field from the shear catalog however receives much less attention. Currently, it is done by locally averaging the shear estimators within some smoothing kernel. The choice of the kernel size can be quite nontrivial: on one hand, it should be large enough to guarantee sufficient background galaxies; on the other hand, it is desirable to keep the kernel size small for a good spatial resolution. The conventional method may also induce systematic errors in the shear field due to the coupling between its fluctuation and the inhomogeneity of the background source galaxies. 

In this work, based on the method of PDF-SYM in \cite{Zhang2017}, we propose the PDF-Folding method as an alternative way of recovering the shear field from the shear catalog. The details of the new method is given in \S\ref{method}. It is significantly different from the conventional weighted-sum methods. The main differences include: 

1. Instead of recovering the local shear value, PDF-Folding reconstruct the coefficients of the orthogonal modes that are used to decompose the shear field as a whole. It therefore does not require any smoothing kernels; 

2. Inhomogeneous distribution of the source galaxies (as well as masks) are automatically taken into account in the method without incurring systematic biases; 

3. The statistical errors on the coefficients of the orthogonal modes can reach the theoretical minimum in general observing conditions, i.e., with inhomogeneous source distribution and realistic shape noise.  

The above points are demonstrated with analytical reasoning (see \S\ref{method} and the Appendixes) as well as numerical examples using the Illutris-1-Dark simulation data (see \S\ref{numerical}). As real world examples, in \S\ref{real}, we apply the PDF-Folding method on the shear catalog of the HSC survey (constructed from the Fourier\_Quad pipeline) to reconstruct the surface density maps around three massive galaxy clusters identified in SDSS. Comparing to the local reconstruction of the shear field, we show that PDF-Folding shows a consistently better performance. It is mainly because the new method is much less susceptive to the local 
inadequacy of source galaxies or outliers in shape noise.       


The main caveat in PDF-Folding is in the calculation of $N_{i (>0)}-N_{-i}$ in \S\ref{method}, in which we expand the PDF of the shear estimators to the first order in shear. It implies that the shear signal is small enough comparing to the dispersion of the shape noise. This condition is not necessarily well satisfied in the neighbourhood of massive galaxy clusters, and therefore there can be small amount of systematic errors. This is though not a serious issue, as the very central regions of galaxy clusters can be simply masked if necessary. Such an expansion in PDF may also fail when the PDF itself has singularities. This has been shown as an advantage in the PDF-SYM method discussed in \cite{Zhang2017}, because it can in principle lead to almost infinite signal-to-noise ratio in shear recovery. In PDF-Folding, however, we find that such singularities in the PDF lead to significant systematic errors in the recovered shear field. It is because the failure of the expansion invalidates the cancellation of the couplings between different orthogonal modes. Nevertheless, this problem is neither a serious concern in practice, as realistic PDF of the shear estimators rarely contains singularities. 

Finally, we have not considered the impact of the shear measurement errors on the recovered shear field. One cause of the error is the uncertainty of the point spread function (PSF), which could depend on the local density of the stars, and therefore induce a variation of the shear field. This topic will be studied in a future work.


\acknowledgments
We thank the referee for very useful suggestions, which significantly improve the writing of the paper. We also thank Jiaqi Wang, Haojie Xu, Ji Yao, Xiaokai Chen, Zhiwei Shao and Zhi Shen for useful discussions on many technical details. This work is supported by the National Key Basic Research and Development Program of China (No.2018YFA0404504), the NSFC grants (11621303, 11890691, 12073017), and the science research grants from China Manned Space Project (No. CMS-CSST-2021-A01). The computations in this paper were run on the $\pi$ 2.0 cluster supported by the Center for High Performance Computing at Shanghai Jiao Tong University.

\bibliography{SF.bib}{}
\bibliographystyle{aasjournal}

\appendix

\section{The Optimal Choice of $\alpha$}

Using the maximum likelyhood method, let us derive the Cramer-Rao Bound on the statistical uncertainties of the coefficients of the shear field. The likelyhood function is simply the PDF of the observed galaxy ellipticities written as:
\begin{align}
P(e)=P_I(e-g(\vx))=P_I(e-\sum_ka_kf_k(\vx)n^{-\alpha}(\vx))
\end{align}
According to the maximum likelihood estimation (MLE), the uncertainty of $a_w$ ($\sigma_{a_w}$) can be written out as:
\begin{eqnarray}
\sigma_{a_w}^{-2}(MLE)&=&-\sum_i\frac{\partial^{2}\ln P(e_i)}{\partial a_w^{2}} = \int d^2\vx\cdot n(\vec{x})\int de P_I^{-1}(e)\left[f_w(\vx)n^{-\alpha}(\vx) P_I'(e)\right]^{2}\\ \nonumber
&=& \int d^2\vx f_w^{2}(\vx)n^{-2\alpha+1}(\vx)\int deP_I^{-1}(e){P_I'}^2(e)
\end{eqnarray}

In the PF method, the value of $a_w$ is estimated by minimizing the $\chi^2$ defined as:
\begin{equation}
\chi^2=\frac{1}{2}\sum_{i(>0)} \frac{(N_i-N_{-i})^2}{\langle(N_i-N_{-i})^2\rangle}
\end{equation}
The denominator in the above formula can be worked out in the following way:
\begin{equation}
\langle(N_{i(>0)}-N_{-i})^2\rangle\approx \langle(\delta N_{i(>0)}-\delta N_{-i})^2\rangle\approx 2\langle(\delta N_{i(>0)})^2\rangle
\end{equation}
in which the $\delta$ sign refers to the deviation from the statistical mean. In this calculation, we have also neglected the influence of the lensing effect, which yields a slightly non-zero mean for $N_{i(>0)}-N_{-i}$. According to eq.(\ref{ni6}), 
\begin{equation}
\label{ni60}
N_{i(>0)}\approx\int d^2\vec{x} \vert f_w(\vec{x}) \vert n^{\alpha}(\vec{x})\int_{u_{i-1}}^{u_i}d\hat{e} P_I(\hat{e})=\int d^2\vec{x} W(\vec{x}) n(\vec{x})\int_{u_{i-1}}^{u_i}d\hat{e} P_I(\hat{e})=\sum_jW(\vec{x}_j)\Delta N(\vec{x}_j)
\end{equation}
in which $W(\vec{x})=\vert f_w(\vec{x}) \vert n^{\alpha-1}(\vec{x})$ is the weight at $\vec{x}$. The last step in the above equation is a discrete version of the integretion.
Assuming that the galaxy number obeys Poisson distribution, we get:
\begin{equation}
\label{ni61}
\langle(\delta N_{i(>0)})^2\rangle=\sum_jW^2(\vec{x}_j)\langle[\delta\Delta N(\vec{x}_j)]^2\rangle=\sum_jW^2(\vec{x}_j)\Delta N(\vec{x}_j)=\int d^2\vec{x} W^2(\vec{x}) n(\vec{x})\int_{u_{i-1}}^{u_i}d\hat{e} P_I(\hat{e})
\end{equation}
Therefore, we have:
\begin{equation}
\langle(N_{i(>0)}-N_{-i})^2\rangle\approx 2\int d^2\vx\cdot f_w^{2}(\vx)n^{2\alpha-1}(\vx)\int_{u_{i-1}}^{u_{i}}d\hat{e}P_I(\hat{e})
\end{equation}
Combining with eq.(\ref{ni50}), we get:
\begin{equation}
\chi^2=(\hat{a}_w-a_w)^2\frac{\left[\int d^2\vx\cdot f_w^{2}(\vx)\right]^2}{\int d^2\vx\cdot f_w^{2}(\vx)n^{2\alpha-1}(\vx)}\sum_{i(>0)} \frac{\left[P_I(u_i)-P_I(u_{i-1})\right]^2}{\int_{u_{i-1}}^{u_{i}}d\hat{e}P_I(\hat{e})}
\end{equation}
In the limit of very small bin size, we can turn the summation in the above formula into an integral form as:
\begin{equation}
\chi^2=\frac{(\hat{a}_w-a_w)^2}{2}\frac{\left[\int d^2\vx\cdot f_w^{2}(\vx)\right]^2}{\int d^2\vx\cdot f_w^{2}(\vx)n^{2\alpha-1}(\vx)}\int deP_I^{-1}(e){P_I'}^2(e) 
\end{equation}
which implies that the error on the parameter $\hat{a}_w$ is:
\begin{equation}
\sigma_{\hat{a}_w}^{-2}(PF)=\frac{\left[\int d^2\vx\cdot f_w^{2}(\vx)\right]^2}{\int d^2\vx\cdot f_w^{2}(\vx)n^{2\alpha-1}(\vx)}\int deP_I^{-1}(e){P_I'}^2(e) 
\end{equation}
According to the Cauchy-Schwarz Inequality, we have:
\begin{equation}
\left[\int d^2\vx\cdot f_w^{2}(\vx)n^{2\alpha-1}(\vx)\right]\cdot\left[\int d^2\vx\cdot f_w^{2}(\vx)n^{1-2\alpha}(\vx)\right]\ge\left[\int d^2\vx\cdot f_w^{2}(\vx)\right]^2,
\end{equation}
therefore, $\sigma_{\hat{a}_w}(PF)\ge\sigma_{a_w}(MLE)$, and the equality is achieved when $\alpha=1/2$. This completes our proof of our statement in \S\ref{orm}.

\section{Reconstruction of the Foreground Surface Density Field}

Let us discuss about reconstructing the surface density distribution $\Sigma(\vec{x},z_l)$ around the foreground lens at redshift $z_l$. The background shear signal is related to the foreground density field via $\kappa(\vec{x},z_s)=\Sigma(\vec{x},z_l)/\Sigma_c(z_l,z_s)$, where the critical surface density $\Sigma_c$ is defined as:
\begin{equation}
\Sigma_c(z_l,z_s)=\frac{c^2}{4\pi G}\frac{D_{A}(z_{s})}{D_{A}(z_{l})D_{A}(z_l,z_{s})(1+z_l)^2}
\end{equation}
Where $D_{A}(z)$ is the angular diameter distance, and the factor $(1+z_l)^2$ is from our use of the comoving scale. To use all the background galaxies to reconstruct the foreground density field $\Sigma(\vec{x},z_l)$, we need to take into account the fact that the background sources are distributed in a range of redshifts. In terms of the shear field, we therefore need to specify at which background redshift we are reconstructing it. Fortunately, under the thin lens assumption, the background shear signal scales with their redshift $z_s$ in a simple enough way through the critical surface density, i.e,  we have:
$g(\vec{x},z_{s1})\Sigma_c(z_l,z_{s1})=g(\vec{x},z_{s2})\Sigma_c(z_l,z_{s2})$. Note that for now we neglect the differences between the shear and the reduced shear. If we choose $z_{s0}$ as our reference background redshift for the reconstructed shear field $g(\vec{x},z_{s0})$, all the background shear at $z_s$ can then be expressed as: $g(\vec{x},z_{s})=g(\vec{x},z_{s0})\Sigma_c(z_l,z_{s0})\Sigma_c^{-1}(z_l,z_{s})$. Following the idea of the PF method, we can expand the reference shear field as $g(\vec{x},z_{s0})\sqrt{n(\vx)}=\sum{a_kf_k(\vx)}$, where $n(\vx)=\int dz\cdot \phi(\vx,z)$, and $\phi(\vx,z)$ is the galaxy number density in the redshift space at angular position $\vx$. To reconstruct $\hat{a}_w$, we should assume that $\hat{g}(\vec{x},z_{s0})\sqrt{n(\vx)}=\hat{a}_wf_w(\vx)$. For any background galaxy at redshift $z_s$, we therefore have: $\hat{g}(\vec{x},z_{s})=\hat{g}(\vec{x},z_{s0})\Sigma_c(z_l,z_{s0})\Sigma_c^{-1}(z_l,z_{s})$, The PDF for the PF method can be formulated as:
\begin{eqnarray}
\label{ni16}
N_i&=&\int_{i*f_w(\vec{x})\ge 0} d^2\vec{x} \int dz\cdot\phi(\vx,z)\vert f_w(\vec{x}) \vert /\sqrt{n(\vec{x})}\frac{\Sigma_c(z_l,z)}{\Sigma_c(z_l,z_{s0})}\int_{u_{\vert i\vert-1}}^{u_{\vert i\vert}}d\hat{e} P(\hat{e})\\ \nonumber
&+&\int_{i*f_w(\vec{x})<0} d^2\vec{x} \int dz\cdot\phi(\vx,z)  \vert f_w(\vec{x}) \vert/\sqrt{n(\vec{x})}\frac{\Sigma_c(z_l,z)}{\Sigma_c(z_l,z_{s0})}\int_{-u_{\vert i\vert}}^{-u_{\vert i\vert-1}}d\hat{e} P(\hat{e}),
\end{eqnarray}
Note that for each background galaxy, we have given it a weight: $\vert f_w(\vec{x}) \vert /\sqrt{n(\vec{x})}*[\Sigma_c(z_l,z)/\Sigma_c(z_l,z_{s0})]$. The lensed PDF can be expanded as:
\begin{eqnarray}
\label{pi12}
&&P(\hat{e})=P_I(\hat{e}+\hat{g}-g)=P_I\left\{\hat{e}+\left[\hat{a}_wf_w(\vec{x})-\sum_k{a_kf_k(\vec{x})}\right]\frac{1}{\sqrt{n(\vx)}}\frac{\Sigma_c(z_l,z_{s0})}{\Sigma_c(z_l,z)}\right\}\\ \nonumber
&\approx&P_I(\hat{e})+\frac{1}{\sqrt{n(\vx)}}\frac{\Sigma_c(z_l,z_{s0})}{\Sigma_c(z_l,z)}\left[\hat{a}_wf_w(\vec{x})-\sum_k{a_kf_k(\vec{x})}\right]P_I'(\hat{e})
\end{eqnarray}
Note that here we assume that the form of $P_I$ does not depend on the redshift. As a result, we have: 
\begin{equation}
\label{ni15}
N_{i (>0)}-N_{-i}\approx2(\hat{a}_w-a_w)\left[P_I(u_i)-P_I(u_{i-1})\right]\int d^2\vec{x} f_w^2(\vec{x})
\end{equation}
The PF method in this definition should therefore yield an unbiased estimate of the parameter $\hat{a}_w$. Following the procedures similar to those in Appendix A, we can get the error on the parameter $\hat{a}_w$ in the limit of small bin size as:
\begin{equation}
\label{err1}
\sigma_{\hat{a}_w}^{-2}(PF)=\frac{\left[\int d^2\vx\cdot f_w^{2}(\vx)\right]^2}{\int d^2\vx\cdot f_w^{2}(\vx)n^{-1}(\vec{x})\int dz\cdot\phi(\vx,z) \left[\Sigma_c(z_l,z)/\Sigma_c(z_l,z_{s0})\right]^2}\int deP_I^{-1}(e){P_I'}^2(e)
\end{equation}
We can further ask whether this form is statistically optimal by finding out the Cramer-Rao bound in this case as:
\begin{eqnarray}
\label{err2}
\sigma_{a_w}^{-2}(MLE)&=&-\sum_i\frac{\partial^{2}\ln P(e_i)}{\partial a_w^{2}} = \int d^2\vx\int dz\cdot\phi(\vx,z)\int de P_I^{-1}(e)\left[f_w(\vx)\frac{1}{\sqrt{n(\vx)}}\frac{\Sigma_c(z_l,z_{s0})}{\Sigma_c(z_l,z)} P_I'(e)\right]^{2}\\ \nonumber
&=& \int d^2\vx f_w^{2}(\vx)n^{-1}(\vx)\int dz\cdot\phi(\vx,z)\left[\frac{\Sigma_c(z_l,z_{s0})}{\Sigma_c(z_l,z)}\right]^2\int deP_I^{-1}(e){P_I'}^2(e)
\end{eqnarray}
From the results of eq.(\ref{err1}) and (\ref{err2}), and using the Cauchy-Schwarz Inequality again, we can show that $\sigma_{\hat{a}_w}(PF)\ge\sigma_{a_w}(MLE)$. Indeed, in general, the equality cannot be achieved without modifying the weighting scheme. 

To further improve our formalism to approach the Cramer-Rao bound, we can change our definition of $n(\vx)$ as: 
\begin{equation}
\label{nxx}
n(\vx)=\int dz\cdot \phi(\vx,z)\Sigma_c^2(z_l,z_{s0})/\Sigma_c^2(z_l,z).
\end{equation}
The weighting in the definition of $N_i$ should be modified accordingly as:
\begin{eqnarray}
\label{ni17}
N_i&=&\int_{i*f_w(\vec{x})\ge 0} d^2\vec{x} \int dz\cdot\phi(\vx,z)\vert f_w(\vec{x}) \vert /\sqrt{n(\vec{x})}\frac{\Sigma_c(z_l,z_{s0})}{\Sigma_c(z_l,z)}\int_{u_{\vert i\vert-1}}^{u_{\vert i\vert}}d\hat{e} P(\hat{e})\\ \nonumber
&+&\int_{i*f_w(\vec{x})<0} d^2\vec{x} \int dz\cdot\phi(\vx,z)  \vert f_w(\vec{x}) \vert/\sqrt{n(\vec{x})}\frac{\Sigma_c(z_l,z_{s0})}{\Sigma_c(z_l,z)}\int_{-u_{\vert i\vert}}^{-u_{\vert i\vert-1}}d\hat{e} P(\hat{e}),
\end{eqnarray}
Note that subtle difference between eq.(\ref{ni17}) and eq.(\ref{ni16}) regarding the positions of the critical surface densities. Once we adopt this formalism, it is straightforward to show that
\begin{equation}
\sigma_{\hat{a}_w}^{-2}(PF)=\sigma_{a_w}^{-2}(MLE)=\int d^2\vx f_w^{2}(\vx)\int de P_I^{-1}(e){P_I'}^2(e).
\end{equation}

In practice, however, we find that the two types of weighting schemes defined in eq.(\ref{ni17}) and eq.(\ref{ni16}) do not lead to significant differences in the final results. For this reason, our results in \S\ref{real} are produced with the PF version defined in eq.(\ref{ni16}), which takes a more straightforward definition of $n(\vx)$. The step-by-step description of shear field reconstruction is a modified version of \S\ref{SBS}:

\quad 1. Set up a rectangular grid for the shear field, and determine a reference background redshift $z_{s0}$, at which the shear field is reconstructed;

\quad 2. Count the galaxy number density $n(\vec{x})$ in each grid;

\quad 3. Fill up the masked area with galaxies of the average number density and random ellipticities/shear estimators, and assign the reference redshift $z_{s0}$ to all of them;

\quad 4. Determine a set of orthogonal and complete functions, e.g., Fourier series $f_w(x)$,  to parameterize the shear field as: $g(\vec{x},z_{s0})\sqrt{n(\vec{x})}=\sum_w{a_wf_w(\vec{x})}$, where $w$ is the index of each Fourier mode.  
The rest of the steps are about determining the coefficients $a_w$ one-by-one. 

\quad 5. Each $a_w$ is determined by changing its assumed value $\hat{a}_w$, so that the PDF of the whole background galaxy sample is best symmetrized. The pseudo shear signal for a galaxy of redshift $z$ is given by $\hat{g}(\vec{x},z)=\hat{a}_wf_w(\vec{x})/\sqrt{n(\vec{x})}\cdot\Sigma_c(z_l,z_{s0})/\Sigma_c(z_l,z)$ (without summing over all possible values of $w$). $z_l$ is the redshift of the lens. For each galaxy, its weight is $\vert f_w(\vec{x}) \vert /\sqrt{n(\vec{x})}\cdot\Sigma_c(z_l,z)/\Sigma_c(z_l,z_{s0})$. The sign of $\hat{e}(=e-\hat{g})$ needs to be inverted wherever $f_w(\vec{x})<0$;

\quad 6. Repeat step 5 to find the values of all $a_w$. Calculate the resulting shear field (at the reference redshift $z_{s0}$) with $g(\vec{x},z_{s0})=\sum_w{a_wf_w(\vec{x})}/\sqrt{n(\vec{x})}$. The resulting shear field can be converted to the $\kappa$ field through, e.g., the K-S inversion algorithm, and the $\kappa$ field can be further converted to the surface density distribution.

A point to note is about our assumption that the form of the PDF $P_I$ does not depend on the redshift. This is not true in practice. This fact would require us to further change the weightings defined in eq.(\ref{nxx}) and (\ref{ni17}) to take into account the redshift-dependent shape noise, for the purpose of achieving the optimal statistical uncertainty. We may study this issue in a future work. Our current formalism of PF is at least close to the optimal form. It is also not hard to show, e.g., with eq.(\ref{ni15}), that even if $P_I$ depends on redshift, the PF method would not generate systematic biases at the first order of shear.

\end{document}